\documentclass[%
 amsmath,amssymb,
 aps, 
floatfix,
prl,
onecolumn
]{revtex4-2}

\renewcommand{\selectlanguage}[1]{}
\usepackage{graphicx}
\usepackage{dcolumn}
\usepackage{bm}
\usepackage{algpseudocode}
\usepackage{comment}

\usepackage{xcolor}
\usepackage{helvet}

\newcounter{natalg}
\renewcommand{\thenatalg}{\arabic{natalg}}

\newenvironment{natbox}[1]{%
  \par\refstepcounter{natalg}%
  \vskip 10pt%
  \noindent\begin{minipage}{\linewidth}%
  \noindent\rule{\linewidth}{1.2pt}\par\vskip 5pt%
  {\sffamily\small\bfseries Algorithm~\thenatalg\ \textbar\ \mdseries #1\par}%
  \vskip 5pt\noindent\rule{\linewidth}{0.4pt}\par\vskip 7pt%
  \small\sffamily%
}{%
  \par\vskip 7pt\noindent\rule{\linewidth}{0.4pt}%
  \end{minipage}\par\vskip 10pt%
}

\newcounter{natstep}
\newenvironment{natsteps}{%
  \begin{list}{\textcolor{black!55}{\sffamily\footnotesize\arabic{natstep}}}%
    {\usecounter{natstep}%
     \setlength{\leftmargin}{1.7em}\setlength{\labelwidth}{1.0em}%
     \setlength{\labelsep}{0.7em}\setlength{\itemsep}{2pt}%
     \setlength{\parsep}{0pt}\setlength{\topsep}{0pt}%
     \setlength{\listparindent}{0pt}}%
}{\end{list}}

\begin{document}

\preprint{APS/123-QED}

\title{\textbf{Machine learned designs of functional colloidal foldamers}}

\author{Ryan van Mastrigt}
 \affiliation{%
  Gulliver Lab, CNRS UMR 7083, ESPCI Paris, PSL University, 75005 Paris, France
}%
 \email{Contact author: ryan.van-mastrigt@espci.psl.eu; zorana.zeravcic@espci.fr}
\author{Zorana Zeravcic}%
 
\affiliation{%
  Gulliver Lab, CNRS UMR 7083, ESPCI Paris, PSL University, 75005 Paris, France
}%

\date{\today}

\begin{abstract}
A protein's function follows from the structure it adopts, and in some cases which structure that is depends on the pathway taken. In programmable matter the target is fixed before assembly, and whatever else forms is treated as error. Here we show that pathways themselves form a design space. Using reinforcement learning, we fold model DNA-coated droplet chains into rigid two-dimensional geometries, uncovering two classes of pathways: downhill, in which bonds are only added, and detour, in which bonds are broken and remade before the target is reached---for some the only route that exists. Coarse-graining pathways by interactions gives experimentally realizable protocols. Some produce one geometry, others several: structures sharing a detour route can be cycled between, while those that coexist assemble into superstructures inaccessible to a uniform product. Function emerges from the pathways rather than being designed. Designing the process instead of the components could give colloidal materials that reconfigure on demand.
\end{abstract}

\maketitle

\begin{figure*}[t]
    \centering
    \includegraphics[width = 1\linewidth]{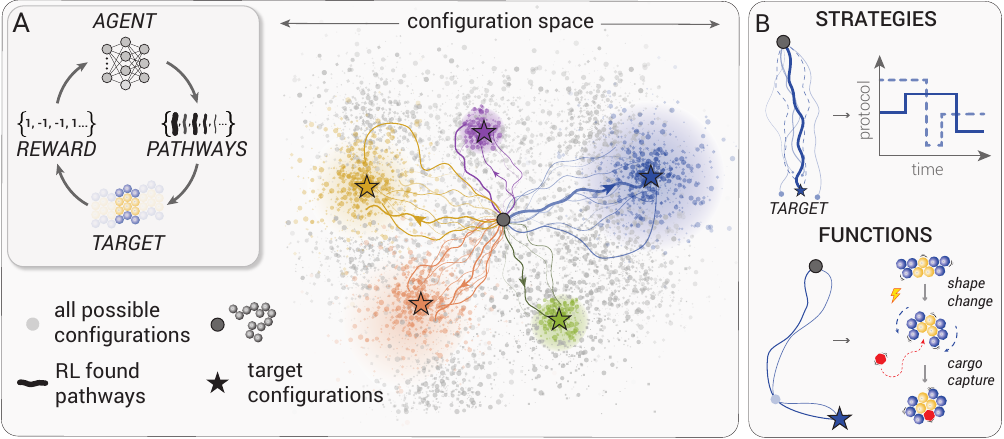}
    \caption{\textbf{Reinforcement learning uncovers folding strategies and emergent functions in colloidal chains.} 
   A. A model-free reinforcement learning agent folds a colloidal chain toward a target structure by switching interactions on and off, rewarded for reaching the target. Lines are the folding pathways it finds---ordered sequences of configurations visited from the open chain (gray circle, center) outward---drawn on a two-dimensional sketch of the high-dimensional configuration space; stars mark the targets the agent is trained on. Repeated test runs follow different pathways, because each starts from a freshly equilibrated chain and folding is stochastic, so a single target is reached by many distinct sequences of interaction switches. Each shaded region groups where those pathways end: a spread around the target rather than the target alone, which is why pathways need not terminate exactly on a star. The extent of each region, and its overlap with others, reflects that spread as much as the underlying landscape topology.
   B. This distribution of pathways is the object of study, and it holds two kinds of insight. Strategies: pathways reaching the same target (blue star) differ in the order in which interactions are switched, and each order translates into a temporal protocol that can be run in experiment (the solid and dashed pathways give the solid and dashed protocols). Functions: configurations visited along the way can do what neither the open chain nor the target can: a ladder configuration (blue circle) reconfigures into a structure with a cavity, that can capture a cargo (red particle). Function can therefore emerge from the folding process itself, without being encoded in the target structure.
    }
    \label{fig1}
\end{figure*}


Folding provides a versatile route to programmable matter: a single strand of DNA~\cite{rothemund2006dnaorigami,Seeman2018DNA}, a chain of emulsion droplets~\cite{mcmullen_freely_2018,mcmullen_self-assembly_2022}, or a template of rigid panels with flexible hinges~\cite{Pinto2024,Neves2026,pandey2011, bircan2020bidirectional-443,Lin2020, coulais_multi-step_2018} can reach a target structure through successive folding steps. Because the components stay connected throughout, the object can be driven from one structure to another without being taken apart~\cite{marras2015programmable,kim2023paperfolding, Pinto2024}. Which structure is reached depends on the path taken~\cite{SternPinson2017, niu2019magnetic-692}. The same holds in biology: in some proteins the native state is not the most stable one, and which fold is reached is set by kinetics rather than by the stability of the endpoint~\cite{Sohl1998,Baker1994}. In other cases external conditions decide which route is followed, like in metamorphic proteins where a change in temperature alone can lead to fold switching~\cite{murzin2008metamorphic-e0a, liwang2025unveiling-0ea}. Design efforts have nevertheless focused on shaping funneled landscapes, with alternative routes typically suppressed. In self-folding sheets, for example, the topology of folding pathways can be shaped by adding imperfections such as stiff creases, designing out the exponentially many misfolding branches \cite{Stern2018,SternPinson2017}; in folding nets the bond specificity between tile edges can be optimized to exclude a competing shell entirely~\cite{Pinto2024}, or combining independent designs with different bond energies makes each structure dominant in its own temperature window, while suppressing all the others~\cite{Bupathy2022}. Here we treat the full set of pathways to a target as the design space (Fig.~\ref{fig1}).

\bigskip

\noindent \textbf{The pathway space.}
We realize this design space in model colloidomers: chains of freely-jointed particles held together by an irreversible backbone, whose folding is driven by secondary (weaker) specific interactions between particle flavors~\cite{mcmullen_self-assembly_2022}. In experiments, these interactions can be switched on and off as the chain folds, by temperature that acts on DNA-bonds between particles~\cite{mcmullen_self-assembly_2022} or by DNA reaction networks~\cite{gines_microscopic_2017,Montagne2011,Montagne2016}. We focus on chains with two flavors, blue (B) and yellow (Y), giving three secondary interactions, BB, BY and YY. For a fixed sequence of flavors, the design variable is then the protocol---the order and timing of interaction switches---rather than the building block. 

This pathway space is spanned by states, which correspond to bond configurations of the chain, and by the transitions between them, defined by the making or breaking of a single bond; a sequence of such transitions defines a pathway. The resulting space is large: even a chain of eight particles can adopt over six thousand bonding topologies connected by over fifty thousand transitions (Supplementary Information). The outcome is also sensitive: a small change in the protocol can change the structures reached entirely, and neither the reachable structures nor the routes between them are easy to know in advance. Enumeration is still possible at this size but it scales poorly with $N$, and methods that optimize protocols over an explicit transition network become impractical beyond it~\cite{Trubiano2022}.

Machine learning has become a natural computational tool for navigating such spaces~\cite{dijkstra2021from-dd3}. It is used for design of building blocks: evolutionary algorithms~\cite{coli2022inverse-009} and differentiable simulation~\cite{Goodrich2021,King2024} tune particle interactions to stabilize target structures. 
In each case the design variable is the building block flavor, or a sequence or the interaction, and the target is an equilibrium property. Machine learning is also used for design of processes: automatic differentiation of kinetic models identifies protocols that steer assembly past kinetic traps~\cite{jhaveri2024discovering-8e5,Engel2023}, and reinforcement learning finds time-dependent protocols that promote crystallization or explore what structures can form~\cite{whitelam_learning_2020,whitelam2021neuroevolutionary-a5f}.

Here we train a model-free reinforcement learning agent to fold colloidomers by switching interactions on and off in response to the chain's current bond configuration, rewarded along the way until the target is reached (Fig.~\ref{fig1}A). Unlike a fixed protocol, this type of exploration takes different routes through configuration space, allowing training to uncover multiple successful pathways to the same structure, each reached by a different sequence of interaction switches. We retain the full ensemble of successful pathways that reach the target (shaded region around each target in Fig.~\ref{fig1}A) and ask what it reveals about reconfigurability. 
Grouped by sequences of switches, these pathways can be translated directly into experimental protocols, so that different routes to the same structure correspond to different folding strategies (Fig.~\ref{fig1}B).  These routes also pass through transient configurations that can acquire functions of their own, e.g., a folding intermediate that briefly opens a cavity can capture cargo and carry it until its release through reconfiguration (Fig.~\ref{fig1}B).

\begin{figure*}[ht!]
    \centering
    \includegraphics[width = 1\linewidth]{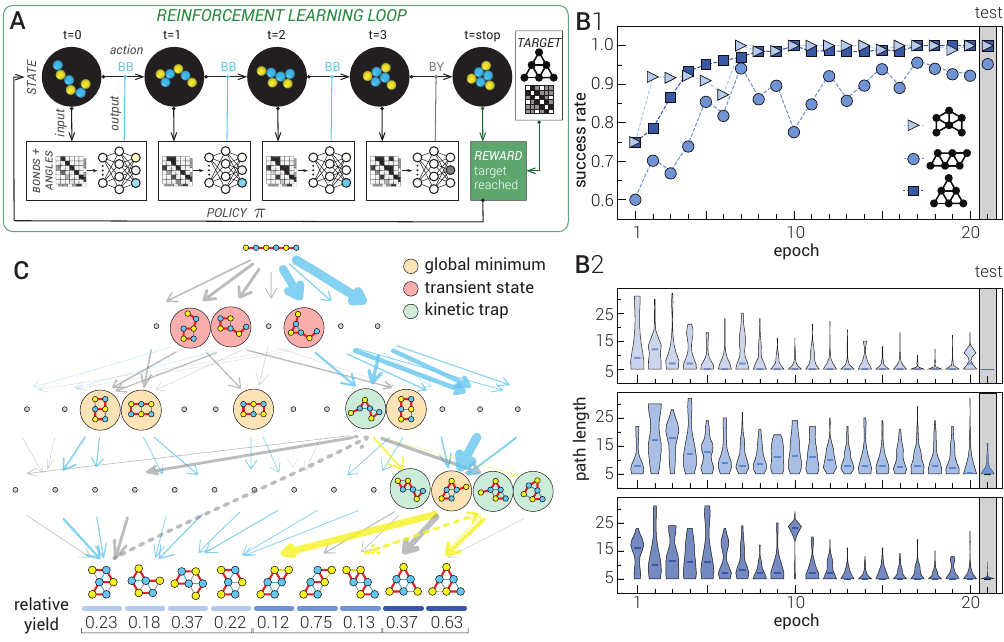}
    \caption{\textbf{Training an RL agent reveals the diversity of successful folding pathways.}
    A. The RL loop. DPD simulations produce chain configurations that are encoded as an adjacency matrix of bonds and interior folding angles. This \textit{state representation}, together with the target, forms the agent's input. The agent outputs a probability distribution over actions, i.e., switching secondary interactions on or off (BB, blue; BY, gray; YY, yellow). The sampled action is passed back to the simulation, which runs until the next folding event (bond making or breaking) defines a new state. An episode ends at a stop condition and a reward is given for reaching the target. Results in this figure focus on the $N=6$ alternating chains, that can fold into three different rigid structures.
    B1. Success rate (the fraction of pathways reaching the target) rises during training for all three target geometries (see legend). At test time, the trained agent reaches the target with near-perfect success (gray band).
    B2. Pathway length (the number of states visited) decreases and its distribution narrows during training, converging for all targets (legend in B1). Violins show the distribution over sampled pathways per epoch and horizontal lines mark medians. 
    C. Folding tree of all discovered pathways, with each row grouping states with the same number of secondary bonds, from unfolded chain (top) to the rigid targets (bottom). Arrow thickness indicates how often a transition is sampled, and color denotes the type of secondary bond formed (solid lines) or broken (dashed lines). Circles classify visited configurations as global minima, transient states or kinetic traps; small gray dots mark sampled configurations where interaction change did not occur. Numbers below each target fold (i.e., each distinct backbone arrangement within a given geometry) represent relative fold yields, normalized separately for each geometry.}
    \label{fig2}
\end{figure*}

\bigskip

\textbf{Exploring folding pathways.} We combine dissipative particle dynamics (DPD) simulations of two-flavor colloidomers with reinforcement learning to explore the space of folding pathways. We represent each state by its secondary bond configuration and backbone orientations (Methods). These features, together with a representation of the target geometry, are passed to an \textit{agent} and a \textit{critic}, implemented as Convolutional Neural Networks (CNNs) mapping states to action probabilities and expected reward, respectively. An action corresponds to switching one secondary interaction on or off. 

Starting from an equilibrated chain and resetting after successful folding or timeout when a predefined maximum number of folding steps is reached, the agent explores state space stochastically and is trained to maximize the expected reward (Fig.~\ref{fig2}A). Because the RL agent learns from sampled pathways, the framework requires no precomputed enumeration of states and transitions, and therefore scales to longer chains where the full landscape becomes intractable to map (we use action masking for $N \leq 8$ as an optional accelerant that is switched off for larger $N$; see Methods).

Training of our RL agent is first done on $N=6$ chains with alternating sequence (BYBYBY), for which the state space is small and downhill strategies are known~\cite{mcmullen_self-assembly_2022}. During training the agent converges to policies that reliably reach the target geometries (Fig.~\ref{fig2}B1), with pathway lengths decreasing (Fig.~\ref{fig2}B2). At test time, by taking the most probable action at each step, the agent reaches the target in most pathways, and the strategies it finds include the known downhill solutions.

To understand how the agent reaches the target, we classify the states at which it switches an interaction (Fig.~\ref{fig2}C). 
A state is a global minimum (GM) when its bond count is the maximum achievable given the interactions that are turned on. If no additional bonds of those types can form and the state is not a GM, it is a kinetic trap (KT). All other states are transient states (TS), from which additional bonds of interaction types that are on can still form. This labeling depends on which interactions are on, so a configuration marked as a kinetic trap under one may be a transient state under another set of interactions.

Most states where an interaction switches are either GMs or KTs (Fig.~\ref{fig2}C). A fixed downhill protocol drives the chain toward a global minimum but stalls whenever it reaches a kinetic trap, limiting its yield. The RL agent circumvents these traps by activating additional interactions or breaking bonds, opening alternative pathways to the target.  
Each of these routes can reach the target with near-unit success. Success here is the fraction of pathways that reach the target for a single chain under a policy that responds to its current state---not the yield of an ensemble under a fixed protocol (which we turn to below). Rather than a single optimal route, the agent reveals an ensemble of successful pathways, raising the question of what distinguishes them.

\begin{figure*}[b]
    \centering
    \includegraphics[width = 1\linewidth]{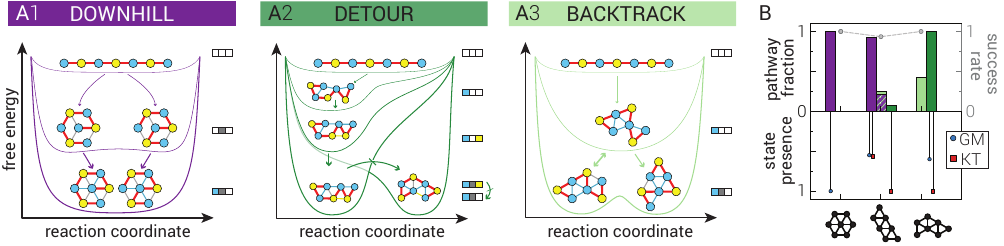}
    \caption{\textbf{Folding pathways classify into distinct strategy types.}
    A. Every successful pathway is either downhill (A1, purple) or detour (A2, dark green). However, both strategies may additionally include some backtracking steps (A3, light green). Within downhill strategies, the chain gains bonds monotonically without ever breaking one. Conversely, within detour strategies states beyond those needed for direct descent are visited: the chain commits to a wrong state, one from which the target cannot be completed, before breaking bonds and refolding toward the target. Backtracking, \textit{i.e.}, breaking a bond to escape a dead-end configuration, can occur within either downhill or detour pathways. In the schematic of the landscapes, boxes on the right indicate which secondary interactions are switched on at each stage of a strategy.
    B. Different targets require different folding strategies: the leftmost target (flower) is reached exclusively by downhill pathways (purple bars), the rightmost (turtle) exclusively by detour pathways (dark green bars), and the middle (ladder) by both classes. Backtracking steps may occur in both strategies (light green bars; hatched purple marks the downhill fraction). Top: fraction of pathways in each class, produced by the \textit{trained agent} for three $N=7$ target geometries (bottom), with success rates given by gray data (right axis). 
    Bottom: fraction of pathways that pass through a global minimum (GM, blue circles) or kinetic traps (KT, red squares) grouped per folding strategy for each target; a single pathway may visit both KT and GM.
    }
    \label{fig3}
\end{figure*}

\bigskip

\noindent \textbf{Folding pathways fall into two classes. }Extending training to alternating chains of $N=7$ and $8$ across all target geometries reveals two classes of successful pathways. We classify each pathway by its length, after reducing it to a self-avoiding path: it either matches the shortest path from the unfolded chain to the target, or exceeds it (Methods). The reduction removes \textit{backtracking}, in which the chain folds into a dead end, breaks a bond to return to a previously visited state, and resumes its route (Fig.~\ref{fig3}A3). Since path length is discrete, the distinction is exact and requires no threshold. The first class we find is \textit{downhill} folding, in which bonds are only added (Fig.~\ref{fig3}A1), while the second is \textit{detour} folding, in which the chain first reaches a state from which the target cannot be completed, breaks bonds, and refolds into the target (Fig.~\ref{fig3}A2). Downhill and detour are therefore mutually exclusive, while  backtracking can occur within either class. The target geometry determines the mixture of successful strategies the RL agent finds. Some targets are reached only by downhill pathways in our sampled ensemble, some only by detours, and others by both (Fig.~\ref{fig3}B). We also find that both classes are dominated by the same landmarks: nearly every pathway the agent finds passes through at least one global minimum or kinetic trap, and pathways confined to transient states are rare (Supplementary Information).

For targets with detour only paths, breaking and remaking bonds is not an optimization but a requirement. This sequence of steps echoes chaperone-assisted protein folding, yet arises here in a short chain with only three interaction types. 
Error correction is therefore not something the system has to be given; it follows from the structure of the folding problem. In biology it relies on dedicated machinery evolved to rescue misfolded states~\cite{chakraborty2010chaperonin-catalyzed-67a, priya2013molecular-611,mattoo2014molecular-2f5}, and in synthetic assembly it has been built into the components, either by letting them sense their environment and change their binding strength~\cite{Zhu2024} or by adding a species that catalytically relaxes trapped intermediates~\cite{Pokhrel2024}. Here the rescue is achieved neither in the components nor by added species, but through the interaction protocol, and can therefore be implemented in experiment with the same external switches that drive folding.

\begin{figure*}[t]
    \centering
    \includegraphics{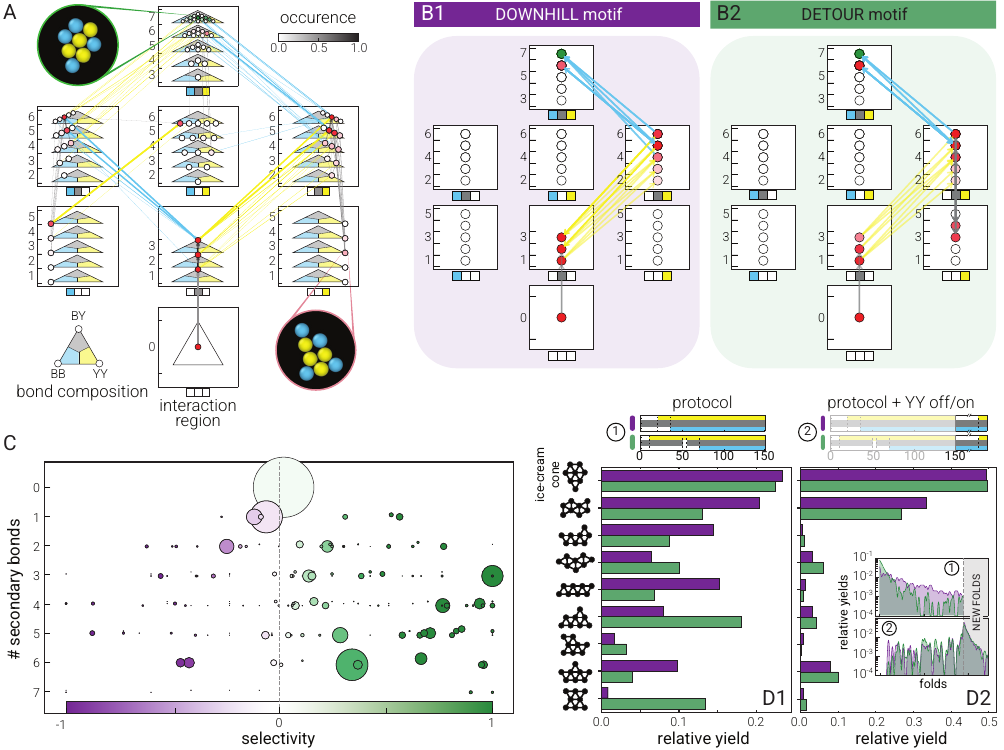}
    \caption{\textbf{Interaction protocols partition accessible structure space.}
    A. Action graph representing all states (circles) and transitions (lines) visited by the trained RL agent for one target ($N=8$ ice-cream cone). States are grouped into eight interaction regions defined by which secondary interactions are active (colored boxes below each group). Within each region, states are arranged by bond count (rows) and bond composition (barycentric coordinates, key in bottom left). Circle shading indicates how often a state is visited (normalized by the most-visited state); green marks target configurations, red marks all others. Arrows connect states visited in succession, with thickness indicating transition frequency, and color the interaction switched on (upward) or off (downward).
    B. The action graph decomposes into motifs, i.e., connected subgraphs that each correspond to a temporal protocol. A downhill motif (B1, purple) contains only transitions that add interactions (with backtracking permitted).  A detour motif (B2, green) includes a bond-breaking transition into a different interaction regime before refolding. These two motifs account for 55\% of sampled pathways.
    C. The two protocols in D1 explore largely non-overlapping regions of state space. States are grouped by canonical adjacency matrix and scored by selectivity, from exclusively downhill (-1) to exclusively detour (+1); marker area scales with total visitation (gathered from 1000 single chain runs per motif).
    D. Each motif translates into a temporal protocol, termed \textit{environment} because it sets the external conditions under which folding occurs (top bars). Bulk simulations of 50 chains per box, ten independent runs per environment, give the relative yield of each of the nine $N=8$ geometries (icons at left). D1. The two environments produce different yield distributions---the same chains, folded under different protocols, reach final states in different ratio. D2. The same environments with an YY off/on motif appended  (Supplemental Information). This motif is not imposed, it appears on its own in three of the ten motifs isolated from the action graph. Although the two environments visit largely disjoint regions of state space (C) and select different backbones, appending the same second motif drives both to the same distribution over geometries. The two panels are two operating modes of the same chain: selectivity by protocol in D1, and robustness to protocol in D2. The reason can be seen in the insets: in the perturbed environment (YY off) gives the partially folded chains access to additional backbone arrangements of dominant geometries they can settle into after the perturbation is over (YY on).
    }
    \label{fig4}
\end{figure*}

\bigskip

\textbf{Mapping strategies to protocols.} The ensemble of discovered pathways can be reduced to a small set of experimentally realizable protocols by coarse-graining them into \textit{action graphs} (Fig.~\ref{fig4}A, Methods). These follow the graph representations developed for multistable materials driven by several independently switchable inputs~\cite{meulblok2026path}, where grouping states by which inputs are currently active makes an otherwise intractable combinatorial response legible, and scales to many inputs. Our switchable inputs are the three secondary interactions, so, in an action graph, all states sharing the same set of active interactions form a single interaction region (squares in Fig.~\ref{fig4}A). The system can explore states within a region without changing the external conditions, while transitions between regions require switching an interaction on or off.

We use action graphs on the example of an $N=8$ alternating chain folding into the ice-cream cone geometry (Fig.~\ref{fig4}A), which the agent reaches by both downhill and detour pathways. Its full action graph has several outgoing transitions per interaction region and therefore cannot be realized by a single protocol. 
To decompose it, we define \textit{motifs} as distinct sequences of interaction switches, obtained by collapsing repeated switches of the same interaction type (Methods). Two prevalent motifs account for $55\%$ of the sampled pathways and correspond to two folding strategies shown in Fig.~\ref{fig4}B1 and B2. The downhill motif adds interactions monotonically: BY, then YY, then BB, while the detour motif switches BY on, YY on, BY off and on again, then BB, deliberately breaking bonds mid-route. Each motif translates directly into a temporal protocol by introducing every new interaction at the most probable time for a chain to reach the number of secondary bonds at which that switch applies (Methods). We term these protocols \textit{environments} as they set the external conditions under which folding occurs.

Implemented in DPD simulations, the two environments drive the same chain to different distributions over all rigid $N=8$ geometries (Fig.~\ref{fig4}D1 and the inset (1) in Fig.~\ref{fig4}D2). The environments select different pathways: defining a per-state selectivity $(V_{\mathrm{detour}}-V_{\mathrm{downhill}})/(V_{\mathrm{detour}}+V_{\mathrm{downhill}})$, where $V_x$ is the visitation count under environment $x$, we find that the chains explore largely non-overlapping regions of state space (Fig.~\ref{fig4}C, Supplementary Video~1).

Appending an additional motif, an YY off/on step, to either environment changes this. Both distributions shift toward the same two geometries and the difference between environments largely disappears (Fig.~\ref{fig4}D2, Supplementary Video~1). The two environments select mostly different backbones, so the same YY off/on step acts on populations with little in common (Supplementary Information). The insets show what happens: with YY off, partially folded chains gain access to backbone arrangements of the dominant geometries that were unreachable under either environment alone, and settle into them when YY returns. We identify two mechanisms underlying this shift: isomerization, when one backbone arrangement switches to another while maintaining the geometry; and what we term ``repair'', when other geometries refold into the target one (Supplemental Information). Protocol choice therefore offers different modes to this chain: without YY off/on the environment selects the outcome, and with it the outcome is robust to the environment. This behavior, however, is not generic. Repeating the same off/on step on BY or on BB leaves the two distributions distinct (Supplementary Information). What determines the outcome is therefore not perturbation as such, but which specific interaction is cycled for.

\bigskip

\begin{figure*}[t]
    \centering
    \includegraphics{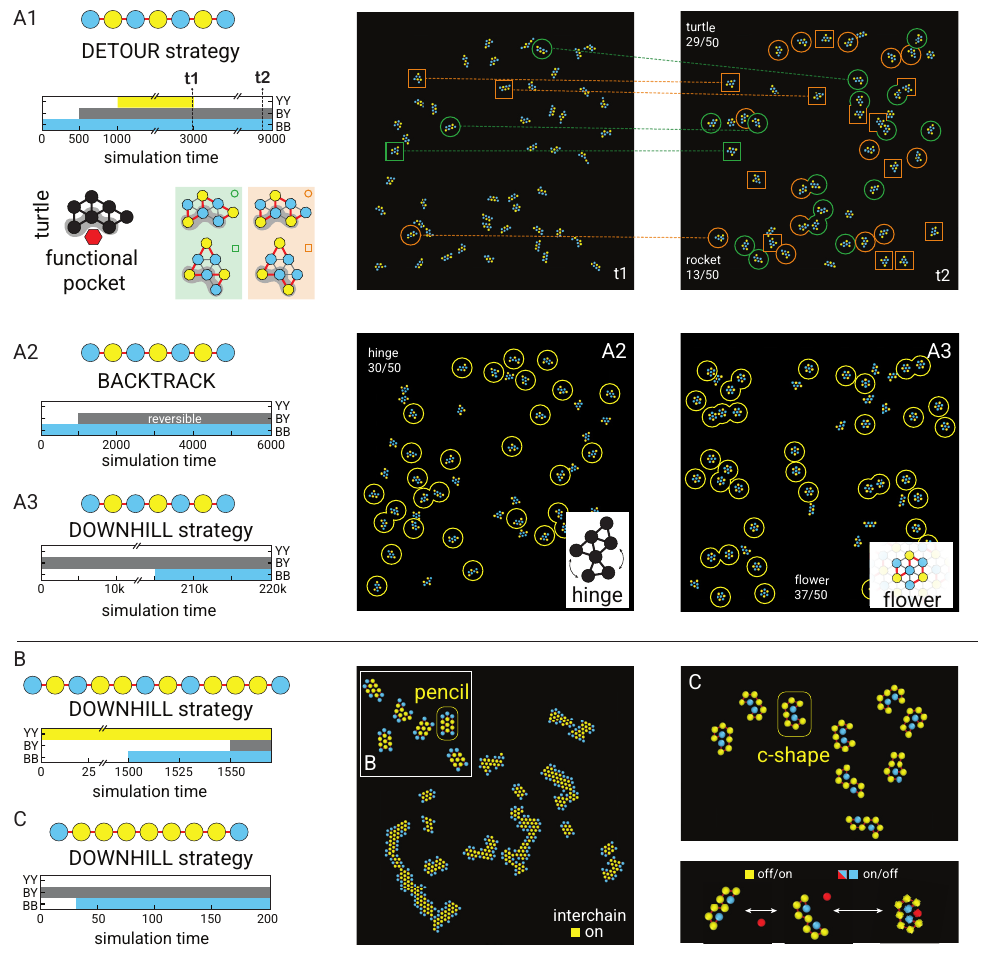}
    \caption{\textbf{Interaction protocols select function, not only structure.} All panels in A use the same alternating $N=7$ chain but different protocols (colored bars left of snapshots). Simulation snapshots are representative of a box of 50 chains.
    A1. A detour strategy found by the agent for the turtle geometry, in which YY is switched on and off again. In the left we show snapshots before (t1) and after (t2) the detour step; dashed lines track representative examples between them. The protocol yields many turtles (marked by circles, 29/50), but the same box also contains rockets (marked by squares, 13/50). Both carry a binding ``pocket'' of the same two compositions, YBB (green) and YBY (orange), shown below the protocol (red particle indicates a cargo the pocket could accept). The functional feature is shared across two geometries and was not specified in the target.
    A2. Under a protocol that includes backtracking, if BY is held weak (``reversible''), BY bonds continue to break and reform. The resulting structures retain a floppy mode (yellow circles, 30/50) and act as hinges.
    A3. Under a downhill strategy the chain folds into the flower (inset), a single rigid geometry with no pocket and no floppy mode, that could be used to tile the plane (yellow circles, 37/50).
    The RL agent works on arbitrary sequences and on higher $N$. B. A $N=12$ chain with a given sequence is folded under a downhill strategy into pencil shapes, which then associate into extended ribbons when interchain YY interactions are switched on. As other structures also form next to the pencil shape, they also participate in supracolloidal assembly, contributing with divots along the ribbons. C. A $N=9$ sequence folded into C-shaped and related structures. The pathways to the shape uncover a hinged intermediate through which the C-shape structure can open and close. The sequence of simulation snapshots below show capture and release of a red particle over time, through cyclic switching of interactions.
    }
    \label{fig5}
\end{figure*}

\noindent\textbf{From fold to function.} So far we have used alternating chains, and Fig.~\ref{fig5}A still does, showing folding at $N=7$. Panels B and C use designed sequences from refs.~\cite{mcmullen_self-assembly_2022, blot2026thesis} at $N=12$ and $N=9$, demonstrating that the approach depends on neither the sequence nor the chain length.

The same chain, under three protocols, gives three different kinds of object (Fig.~\ref{fig5}A1-A3). Asked for the turtle geometry, one of the $N=7$ geometries that is not a foldamer (does not have a downhill solution for the alternating two-flavor sequence~\cite{mcmullen_self-assembly_2022}), the agent returns a detour strategy that reaches it reliably, mostly by first adopting the ladder geometry and then refolding (yield $> 50\%$, without any optimization). Some of the chains under the same protocol fold into rockets, the geometry the agent was not tasked to reach. However, both geometries have something in common, a ``binding pocket'' of the same compositions (Fig.~\ref{fig5}A1, Supplementary Video~2). The pocket arises from pathways rather than the geometry of the target. If we instead immerse the chains into a different environment, one coming from the backtracking strategy the agent found, the structures reached develop hinges, and at the two limits of the hinge motion they close into the turtle and rocket geometries (Fig.~\ref{fig5}A2, Supplementary Video~3). A third protocol, coming from a downhill strategy, leads to chains adopting a rigid state: a flower geometry with checker-board pattern of B and Y particles, that can tile a plane (Fig.~\ref{fig5}A3, Supplementary Video~4).

The choice of environment propagates beyond the single chain. An $N=12$ sequence folds into pencil structures that associate into ribbons when interchain YY is switched on (Fig.~\ref{fig5}B, Supplementary Videos~5 and 6). Unlike the uniform ribbons from a monodisperse foldamer population (Supplementary Information), these carry hinges and divots along their edges, contributed by the minority folds the protocol also produces. An $N=9$ sequence folds into C-shaped structure with an open cavity whose pathways pass through a hinged intermediate. The chain can thus be cycled between a state that encloses a cargo particle and an inert extended state, with capture-and-release set entirely by the external protocol and no change to the chain itself (Fig.~\ref{fig5}C, Supplementary Videos~7 and 8).

\bigskip

\noindent\textbf{Conclusion and discussion.}
Function is increasingly described as a property of an ensemble rather than of a single structure. In proteins this has been proposed as an extension of the sequence-structure-function paradigm, in which sequence encodes a landscape of accessible states whose populations are redistributed by cellular context~\cite{Tripathi2026}. Computationally, the same relation has been inverted for design, tuning a sequence until its conformational ensemble meets a target, including targets defined by the response to salt or temperature~\cite{krueger_shrinivas_2025}. In the design case the ensemble is an equilibrium object, and far-from-equilibrium properties fall outside the framework. 
History-dependent behavior is recognized in proteins---hysteresis, kinetic trapping, long-lived substates---but as a property to be described rather than a variable to design with. In this work the sequence is fixed and the protocol is the design variable, and the ensemble is one of pathways rather than of structures.

Our reinforcement learning agent finds such pathways without prior knowledge of the folding landscape.
For some target geometries no downhill route exists for a given sequence~\cite{mcmullen_self-assembly_2022}, and the agent reaches them through detour pathways. The capacity to break and remake bonds is, therefore, a precondition for accessing these folds, and any system that must reach them---through evolved machinery or through external switches---is under pressure to acquire it. Biology met this pressure with chaperones~\cite{hartl2011molecular,thirumalai2020iterative, Finka2016} and ATP-driven conformational cycling~\cite{Goloubinoff2018}; our chains meet it with the same switches that drive folding.

The system thus realizes a colloidal analogue of Waddington's landscape: the space of structures is fixed by the chain's flavor sequence, just as the space of developmental fates is shaped by the genotype \cite{waddington1957strategy,ferrell2012bistability}. Switching interactions reshapes the accessible portion of the landscape, playing the role of a developmental signal that opens some valleys and closes others. Turning an interaction on earlier rather than later does not alter what is possible in principle, only which region remains reachable, and thus which fold the chain is guided into.

\bigskip 
This framing suggests a different design philosophy for self-assembly. Optimizing a protocol for a target yields just that target; designing the pathways instead returns a set of assembly strategies and reconfigurations that one chain can be switched between. Multifarious mixtures achieve breadth in a different way, storing many structures in a large repertoire of flavors and retrieving one by biasing which nucleates~\cite{Murugan2015, evans2024pattern-673}, or by a temperature protocol that selects between two structures designed into the mixture~\cite{Bupathy2022}. Here the repertoire is only two flavors and one connected object, and the selection is made by the order of switches.

For materials design, the consequence is that the design variable shifts from the building block to the protocol. One chain, run under different environments, yields different folds, superstructures and reconfiguration capabilities, delivering the promise of programmable matter we set out with: a common set of components directed into many structures and functions on demand, without redesigning the components themselves. Such reconfigurable colloidal materials open routes to colloidal robotics~\cite{liu_colloidal_2023, chen2024materials-c1a}, to machines that capture and release cargo~\cite{sacanna2010lock-686, zeravcic_colloquium_2017}, and to computation embedded in matter~\cite{yasuda2021mechanical-dd3}. Extending the approach to three dimensions, to longer chains where the model-free nature of the agent becomes essential, and to experimental implementation are natural next steps toward materials that build, reconfigure, and repair themselves on demand.

\bigskip

\noindent\textbf {Acknowledgements.} We would like to thank Martin van Hecke, Natalie Blot, Maitane Muñoz-Basagoiti, Muhittin Mungan, Pepijn Moerman, Guillaume Gines and Jasna Brujić for useful discussions, and Jasna Brujić and Maitane Muñoz-Basagoiti for feedback on the manuscript. Z.Z. acknowledges financial support from the City of Paris EMERGENCE(S) grant and ANR Tocata grant ANR-22-CE06-0037; R.v.M. acknowledges this publication is part of the project ReFoldamers: Guiding Colloidal Refolding using Machine Learning with file number 019.241EN.008 of the research programme Rubicon which is financed by the Dutch Research Council (NWO).
Simulations were performed using computational resources provided by the Swift cluster of the Gulliver Lab. The authors acknowledge the use of AI tools
(Claude, ChatGPT) to assist with code and manuscript editing. The authors are solely responsible for conceptualization and verification, and assume full responsibility for all contents of the present manuscript.

\bigskip

\noindent{\bf Author Contributions:}  
R.v.M. and Z.Z. conceived and performed the study, and wrote the manuscript. R.v.M. wrote all the codes, ran all the ML training and DPD simulations, and analyzed data.

\bigskip

\noindent{\bf Competing Interests:} 
 The authors declare no competing interests. 

 \bigskip
 
\noindent{\bf Supplementary Information:} 
Supplementary Information is available for this paper.

\bigskip

\noindent{\bf Correspondence and requests for materials} should be addressed to Ryan van Mastrigt (ryan.van-mastrigt@espci.psl.eu) or Zorana Zeravcic (zorana.zeravcic@espci.fr).

\bibliography{apssamp.bib}

\onecolumngrid
\newpage
\section*{Methods}
Here, we describe the methods we use to simulate our chains, to find viable pathways using reinforcement learning, to analyze those pathways, and to turn them into experimental protocols. 

\subsection{Simulating folding chains}
To simulate a chain of DNA-coated droplets, we use dissipative particle dynamics (DPD). We distinguish between two main types of particles: solvent and colloid. Solvent particles are regular DPD particles~\cite{frenkel2023understanding} subject to pairwise conservative, dissipative and random forces with length scale $\sigma$, mass $m$, cutoff distance $r_c^{\mathrm{DPD}}$, friction coefficient $\gamma$ and temperature $T$. Colloid particles are also subject to pairwise DPD forces between colloid and solvent particles. Additionally, the DNA-mediated interactions between colloid particles are modeled via a pairwise short-range Lennard-Jones (LJ) 96-48 potential:
\begin{equation}
    V_{I} = \begin{cases} 
    4 \epsilon_{I} \left[ \left(\frac{\sigma}{r}\right)^{96} - \left(\frac{\sigma}{r}\right)^{48} \right], & \mathrm{if} \, r\leq r^{\mathrm{LJ}}_{c} \\
    0, & \mathrm{if}\, r > r^{\mathrm{LJ}}_{c}
    \end{cases}
    \label{eq:VLJ}
\end{equation}
where $\epsilon_{I}$ is the interaction strength for interaction type $I$, $\sigma$ is the diameter of the colloidal particles, $r$ is the Euclidean distance between two centers of colloidal particles and $r^{\mathrm{LJ}}_c$ is the cutoff distance. We distinguish between backbone interactions $I={b}$ which are always on, and secondary interactions $I={x y}$ between colloid particles of type $x$ and $y$, which can be switched on or off at will both within the chain (intrachain) and to other chains of colloids (interchain). Pairs of colloid particles for which secondary interactions are turned on are not subject to conservative DPD forces, because the LJ potential already provides a sufficiently strong repulsive force to prevent overlap for $r<r_{\mathrm{min}}$, where $r_{\mathrm{min}}=2^{1 / 48}$. They are always subject to intracolloidal pairwise dissipative and random forces. When training and testing the RL agent, the secondary interaction strengths is always $\epsilon_{I}=80$ for all interaction types $I$ that are switched on. This interaction strength is sufficiently strong such that interactions are effectively irreversible. For simulations of chains folding subject to a specific protocol, interaction strengths may be chosen to be reversible for part of the protocol: this corresponds to an interaction strength range of $\epsilon_{I}=[4, 12]$.

The particles are simulated in a two dimensional box of size $L \times L$ with periodic boundary conditions. For simulations with a single chain, the box size $L =N+2$. For bulk simulations with 10 chains $L=25$, for 50 chains $L=75$. The units of simulations are chosen such that $\sigma,\,m,\,T=1$, see Tab.~\ref{tab:DPDparams} for all simulation parameters. We uniformly distribute $N_s=\lfloor\rho L^2 \rfloor$ solvent particles at density $\rho$. Likewise, colloidal chains are initialized by drawing initial positions and angles from an uniform distribution, and placing monomers in a line starting from the initial position in the direction of the angle each spaced $r_{min}$ apart. Initial velocities are drawn from the Maxwell-Boltzmann distribution. Solvent particles overlapping with any colloid particles are randomly displaced until there are no more overlapping solvent-colloid pairs to prevent large repulsive conservative forces. The system is equilibrated for $t_{eq}$ units of simulation time before folding simulations start.

The equations of motion are integrated in a ``kick-drift-kick'' leapfrog algorithm:

\begin{align}
    \mathbf{v}_{i}(t+1/2\Delta t) &= \mathbf{v}_{i}(t) + \frac{\Delta t}{2m} \mathbf{F}_{i}(\{\mathbf{r}(t)\}, \{\mathbf{v}(t)\}), \label{eq:LFv1}\\
    \mathbf{r}_{i}(t+\Delta t) &= \mathbf{r}_{i}(t) + \Delta t \mathbf{v}_{i}(t+1/2\Delta t), \label{eq:LFr} \\
    \mathbf{v}_{i}(t+\Delta t) &= \mathbf{v}_{i}(t+1/2 \Delta t) + \frac{\Delta t}{2 m} \mathbf{F}_{i}(\{\mathbf{r}(t+\Delta t)\}, \{\mathbf{v}(t+1/2\Delta t)\}), \label{eq:LFv2}
\end{align}

where $\Delta t$ is the integration time step and $\mathbf{F}$ is the total force. Because the LJ potential is very short ranged, we integrate the DPD and LJ forces on different time scales: $\Delta t_{s}=10^{-2}$ for the DPD forces and $\Delta t_{c}=10^{-4}$ for the LJ forces. The algorithm is described in Alg.~\ref{alg:mts}. 

\begin{natbox}{Multiple-time-step leapfrog integration over one solvent step $\Delta t_s$.}
\label{alg:mts}
\begin{natsteps}
  \item Compute DPD forces $\mathbf{F}_{\mathrm{DPD}}(\{\mathbf{r}\},\{\mathbf{v}\})$.
  \item Update all velocities $\{\mathbf{v}\}$ [Eq.~\eqref{eq:LFv1}].
  \item Repeat 100 times: leapfrog integrate the intracolloidal DPD and
        Lennard-Jones forces with time step $\Delta t_{c}$
        [Eqs.~\eqref{eq:LFv1}--\eqref{eq:LFv2}].
  \item Update solvent positions $\{\mathbf{r}^{s}\}$ [Eq.~\eqref{eq:LFr}].
  \item Recompute DPD forces $\mathbf{F}_{\mathrm{DPD}}(\{\mathbf{r}\},\{\mathbf{v}\})$.
  \item Update velocities $\{\mathbf{v}\}$ [Eq.~\eqref{eq:LFv2}].
\end{natsteps}
\end{natbox}

\begin{table}[b]
\caption{DPD simulation parameters.}
    \centering
    \begin{tabular}{l|c|c|c|c|c|c|c|c|c|c|c|c|c|c|c|c}
        \textbf{quantity} & $\sigma$ & $m$ & $T$ & $r_c^{\mathrm{DPD}}$ & $r_c^{\mathrm{LJ}}$ & $ \Delta t_s$ & $\Delta t_c$ & $\epsilon_{b}$ & $\epsilon_{I\neq b}$ & $\rho$ & $L$ & $a_{cc}$ & $a_{cs}$ & $a_{ss}$ & $ \gamma$ & $t_{eq}$ \\\hline
        \textbf{value} & $1$ & $1$ & $1$ & $1.1$ & $1.3$ & $10^{-2}$ & $10^{-4}$ & $80$ & $$[0,80]$$ & $3$ & $[N+2, 25, 75]$ & $100$ & $25$ & $25$ & $\frac{1}{2}$ & 10
    \end{tabular}
    
    \label{tab:DPDparams}
\end{table}

\subsection{Reinforcement Learning}
We utilize a model-free reinforcement learning (RL) approach that uses DPD simulations to train on-the-fly and find strategies for folding. Specifically, we use proximal policy optimization (PPO), because it conditionally samples the configuration space in distinct learning episodes, allowing the RL to gradually find and improve upon rare successful folding pathways. We use convolutional neural networks (CNNs) to represent our policy and value functions, because neural networks generalize to new states and the state space is large. 
The policy function is a probability distribution $p(a|S)$ conditioned on state $S$ for action $a$: switching on or off a secondary interaction. The value function $V(S)$ is the expected reward-to-go $\bar{R}(S)$ from state $S$ to the end state following the policy.
The policy network is trained to approximate the probability distribution that maximizes the expected reward-to-go, while the value network is trained to best predict the expected reward-to-go per state following the optimal policy.

The goal of the RL agent is to reliably fold from chain into the desired structure(s). To find folding strategies, our networks interact with our DPD simulations. First, we divide the folding process into discrete steps by halting the simulation after each folding event: a secondary bond formation or breaking. Then, at each step $s$, the policy network is fed a representation of the state of our chain $S_s$ (see Fig.~\ref{fig2}A) and of the desired target geometry. In return the policy network
outputs a probability distribution over the actions $\mathbf{a}$ it can take. 
Finally, we sample an action from this distribution, switch the corresponding interaction on or off in the simulation and continue to run the simulation until the next step. We start from a chain and end the simulation when we reach an end state, defined as either the target state, states for which there are no (valid) actions to take or when the run has reached a predefined maximum number of steps. After an end state, the next step starts again from a newly initialized chain. A learning episode consists of a fixed number of steps specified beforehand.

To update the network parameters, we require a reward to determine how good the traversed paths are. We provide a reward $R_s$ at every step $s$. Specifically, we provide a reward of 1 if the simulation reaches a desired target state $S_T$. For every step taken, we deduct 0.01 to incentivize shorter paths. For states for which there are no valid actions to take, we deduct 0.25. We also provide intermediate rewards: if a folding event created a new bond that matches any of the desired states we add 0.01 points. To prevent greedy policies, the networks are trained on the discounted reward-to-go $\bar{R}_s$: the sum of rewards from step $s$ until the first following end state at step $F$ reweighted by discount factor $\gamma_{\mathrm{df}}$. We note that there was no need for us to use this discount factor, instead we keep it fixed at $\gamma_{\mathrm{df}}=1$. The network parameters update after generating all steps in the episode using gradient-based optimizer Adam~\cite{kingma2014adam}.

The convergence of RL can be improved by masking invalid actions. Preventing the policy network from predicting actions which we know beforehand will not result in a change of state helps the network optimize over the relevant actions faster. We mask the actions which turns off a secondary interaction for states where there are no bonds of that interaction type. Additionally, we mask actions that turn on a secondary interaction when there are no more bonds of that interaction type that can be made. Checking for this requires checking the compatibility between the current state's adjacency matrix and backbone arrangement to all possible rigid states (see the section on masking actions and identifying kinetic traps). If not all rigid states are known \textit{a priori}, this masking can be omitted to ensure the RL is able to explore the full state space. Conversely, this masking can be used to force the RL to explore a smaller state space directly related to a group of specified rigid states.

\subsubsection{State representation and network architectures}
To provide a map from state to action and value, we represent the state $S_t$ as a $N\times N\times (N_T+2)$ tensor with $N$ the number of monomers in the chain and $N_T$ the number of independent target geometries the RL is being trained for. In analogy to digital images, we consider the third dimension of the tensor the channel dimension. Instead of colors, these channels now each represent different information about the state and intended target. The first channel represents the current state of the chain: the upper triangular is the upper triangular part of the adjacency matrix; the diagonal is the sequence of monomer types; the second lower off-diagonal are the internal angles $\phi_{i-1, i, i+1}$ between subsequent monomers $i-1$, $i$ and $i+1$. The second channel represents the target geometry: all target folds' adjacency matrices are summed and normalized by the number of target folds. The remaining $N_T$ channels are used for a one-hot encoding to indicate the intended target in which one full $N\times N$ channel is all ones and the others are all zeroes.

Given this representation of states, convolutional neural networks (CNNs) are a natural choice of architecture. Specifically, our networks are relatively small, consisting of two valid convolutional layers of 16 $1\times 1$ and 32 $3\times 3$ filters. After each convolution, a $\tanh$ activation function is applied. The last convolutional layer is fully connected to a hidden layer of 128 neurons and $\tanh$ activation functions. In the policy network, this hidden layer is in turn fully connected to an output layer of $2N_{I}$ neurons, where $N_{I}$ is the number of secondary interactions. A softmax activation function is applied to the output neurons so that each output neuron corresponds to a probability to either turn on or turn off a interaction type $I$. In the value network, the hidden layer is fully connected to a single output neuron without an activation function. We initialize our networks' parameters $\theta$ using orthogonal Glorot initialization~\cite{glorot2010understanding}.

\subsubsection{Training}
To train the RL, we first collect trajectories of states, actions and rewards by sampling folding in our DPD simulations until a fixed number of steps is reached. If the model is trained for multiple targets at the same time, a new target is chosen at random at the start of a new trajectory. Using these trajectories, we determine the reward-to-go at each step $s$ in the trajectory:
\begin{equation}
    \bar{R}_s = \sum_{s'=s}^{F} \gamma_{\mathrm{df}}^{s'-s}R_{s'},
\end{equation}
where $\gamma_{\mathrm{df}}$ is the discount factor for future rewards and $F$ is the first following end state with $F\geq s$. Note that any state for which the next state is reset to a chain is labeled an end state. To determine if an action was beneficial to reach our goal, we calculate the generalized advantage estimation at each step $s$:
\begin{equation}
    \hat{A}_s =\sum_{s'=s}^{F}(\gamma_{\mathrm{df}} \lambda)^{s'-s} \delta_{s'}
\end{equation}
where $\lambda_{\mathrm{GAE}}$ controls the weighting between bias and variance, and $\delta_s$ is the temporal difference error of the estimated advantage:
\begin{equation}
    \delta_s = R_s + \gamma_{\mathrm{df}} V(S_{s+1}) - V(S_s).
\end{equation}

The reward-to-go and generalized advantage estimation allow us to update our value and policy network parameters. To update the value network parameters $\theta_v$, we minimize the mean squared error between the value function and the reward-to-go:
\begin{equation}
    \mathcal{L}_{v} = \frac{1}{N_B}\sum_{i\in  B} \left( V_{\theta_v}(S_i) - \bar{R}_i \right)^2,
\end{equation}
for the steps $i$ in data batch $B$ of size $N_B$. We update the parameters over randomly selected batches for ten total iterations of the data.

To update the policy network parameters $\theta_p$, we maximize the PPO-clip objective:
\begin{equation}
    \mathcal{L}_p= \frac{1}{N_b}  \sum_{i \in B} \left[ \mathrm{min}\left( \frac{p_{\theta_p}(a_i|S_i)}{p_{\theta'_p}(a_i|S_i)} \hat{A}_i, \mathrm{clip} \left(\frac{p_{\theta_p}(a_i|S_i)}{p_{\theta'_p}(a_i|S_i)}, 1-\xi, 1+\xi \right) \hat{A}_i \right) -\lambda_e \sum_{a} p_{\theta_p}(a|S_i) \ln{p_{\theta_p}(a|S_i)}\right],
\end{equation}
where $a_i$ is the action taken in step $i$, $\theta'_p$ is the set of policy parameters used to collect the trajectories, $\theta_p$ are the parameters to be updated, and $\lambda_e$ sets the weight of the entropy term to promote exploration. The loss function is clipped to prevent the policy function from taking large steps.

We run a new RL step for 100 units of simulation time, and consider a new state to be stable if the adjacency matrix has remained unchanged for 1 unit of simulation time. The parameters we use for training our RL agents are shown in table~\ref{tab:RLparams}. The batch size $N_B$ differs per trained agent, and varies between 128 and 1024.

\begin{table}[b]
    \centering
        \caption{RL parameters}
    \begin{tabular}{l|c | c | c | c | c | c}
        \textbf{quantity} & steps per epoch & learning rate & $\lambda_{\mathrm{GAE}}$  & $\gamma_{\mathrm{df}}$ & $\xi$ & $\lambda_e$ \\\hline
        \textbf{value} & 4096 & 0.0005 & 0.99 & 1.0 & 0.2 & 0.01 
    \end{tabular}

    \label{tab:RLparams}
\end{table}

\subsubsection{Masking actions and identifying kinetic traps \label{subsec:mask}}
To help the RL to converge, we mask actions that do not change the state. That includes actions that turn off interaction types that are not on and actions that turn on interaction types that can not result in the formation of an extra bond of that type. The former is straightforward to check: interaction types $I$ of which there are no bonds formed in the state $S_s$ can not be turned off. This can be checked by comparing the adjacency matrix $A_I$ of all the bonds that can be formed for interaction type $I$ to the adjacency matrix $A_s$ of the state $S_s$. The latter is more involved: it requires checking for kinetic constraints to identify if a state is able to form more bonds of an interaction type.

To check if a state $S_s$ is kinetically constrained and unable to form more bonds of type $I$, we compare state $S_s$ to all (available) rigid states $\Omega=\{S_r\}$. First, we identify for which rigid states the state's adjacency matrix $A_s$ is a subset. These rigid states are likely accessible by $S_s$ via downhill folding. However, $A_s$ does not contain information on the backbone orientation, which can further constrain which rigid states are accessible. To further narrow down the accessible rigid states, we compare the rigid angles $\mathbf{\phi}_s$ of state $S_s$ to the angles of the rigid states. If these angles do not match with a rigid state, the rigid state cannot be accessed from state $S_s$ via downhill folding. If any of the remaining accessible rigid states allow for the formation of more bonds of type $I$ compared to state $S_s$, the action of turning on $I$ is not masked. The full masking action procedure is described in Alg.~\ref{alg:masks}. We note that the masking of actions is not needed to find viable pathways; it only serves to make exploration of pathway space more efficient. We used masking for our RL agent trained on chains of length $N=6$, $7$ and $8$, and not for the agent trained on chains of $N=9$ and $12$.

\begin{natbox}{Computation of action masks.}
\label{alg:masks}
\begin{natsteps}
  \item Once before training, compute the adjacency matrix $A_r$ and the backbone
        angles $\phi_r$ of every rigid target state $S_r \in \Omega$.
  \item Compute the adjacency matrix $A_s$ of the current state $S_s$.
  \item Form the target subset $\Omega' = \{\, S_r \in \Omega : A_s \subseteq A_r \,\}$.
  \item Compute the rigid backbone angles $\bar{\phi}_s$ of state $S_s$.
  \item Form the target subset
        $\Omega'' = \{\, S_r \in \Omega' : \phi_{r} = \bar{\phi}_s \,\}$.
  \item Mask the action that turns interaction $I$ on if
        $A_I \cap A_{r} = \emptyset$ for all $S_r \in \Omega''$.
  \item Mask the action that turns interaction $I$ off if
        $A_I \cap A_s = \emptyset$.
\end{natsteps}
\end{natbox}

\subsubsection{Testing performance}
To test the performance of our RL after training, we conduct a fixed number of test runs. Each test run starts with an equilibrated chain that forms the initial state $S_0$ and input for the policy network. Instead of sampling actions from the probability distribution $p(\mathbf{a}|S_0)$, we instead take the most likely action $a^*=\max_{a^*}p(a^*|S_0)$ and continue the DPD simulation using that action. We repeat these steps until the simulation reaches the target states $S_T$ or exceeds a predetermined maximum number of steps. Test runs are run in sequence for a specific target geometry, each starting with a unique seed for the random number generator.

After generating these test runs, we first check if they are valid: runs are invalid if at any point any of the backbone bonds was broken. The valid runs are prepared for analysis by matching the traversed states to unique labels determined by their adjacency matrices and backbone arrangement. These sequences of labels represents pathways of folding through a discrete folding tree to be used for analysis.

\subsection{Analyzing pathways}
To analyze folding, we map trajectories of our DPD simulations to states in a tree of folding. This allows us to compare and quantify pathways.

\subsubsection{Defining states}
We map a configuration of monomer positions $\{\mathbf{r}\}$ to a state $S$ by adjacency matrix $A$ and interior backbone angles $\phi$. 

The adjacency matrix is calculated by the interparticle distance $r_{ij}$:
\begin{equation}
    A_{ij} = \begin{cases}
        1, & \mathrm{if}\, \left( r_{\mathrm{min}}-\delta< r_{ij}<r_{\mathrm{min}}+\delta \right) \, \wedge \, \left( \exists I \, \mathrm{such}\,\mathrm{that}\, (A_{I})_{ij}=1 \right )\\
        0, & \mathrm{else},
    \end{cases}
\end{equation}
where $\delta$ is a margin of error around the $V_{I}$ minimum. We usually take $\delta=0.05$. $A_I$ is the interaction matrix for interaction $I$. 

The adjacency matrix alone is insufficient to group states: it is unable to distinguish between some kinetic traps and transient states as it contains no information on the orientation of the backbone. To include this information, we calculate the interior angles $\phi_{ijk}$:
\begin{equation}
    \phi_{i,j,k} = \begin{cases}
        \arccos{\left( \frac{\mathbf{r}_{j, i} \cdot \mathbf{r}_{j,k}}{r_{j,i} r_{j,k}} \right)}, & \mathrm{if}\, \mathbf{r}_{j,k} \cdot \mathbf{r}_{j,i}^{\bot} \geq 0 \\
        2 \pi - \arccos{\left( \frac{\mathbf{r}_{j, i} \cdot \mathbf{r}_{j,k}}{r_{j,i} r_{j,k}} \right)}, & \mathrm{else}\, 
    \end{cases}
\end{equation}
where $\mathbf{r}_{i,j} = \mathbf{r}_j - \mathbf{r}_i$, $r_{i,j} = \Vert\mathbf{r}_{i,j}\Vert$, and $\mathbf{r}_{i,j}^{\bot}=\begin{bmatrix}
    0 & 1 \\ -1 & 0
\end{bmatrix} \mathbf{r}_{i,j}$. Note that $\phi_{i,j,k}\in [0, 2\pi]$.

To represent our state, we consider for each bond $(i,j)$ with $i<j$ all the monomers $(i, i+1, ..., j)$ and their backbone angles $b_{ij}=(\phi_{i, i+1, i+2}, ..., \phi_{j-2, j-1, j})$. We then define the matrix $S$ with elements
\begin{equation}
    S_{ij} = \begin{cases}
        1, & \mathrm{if}\, A_{ij}=1 \, \wedge \, i<j \\
        \mathrm{round}(3 b_{ji} / \pi ) - 3 , & \mathrm{if}\, A_{ji}=1 \, \wedge \, i=j+2 \\
        \mathrm{sgn}\left(\sum_a (b_{ji})_a \right), & \mathrm{if} \, A_{ji}=1 \wedge i > j+2 \\
        0 , & \mathrm{else},
    \end{cases}
\end{equation}
where $\mathrm{round}$ and $\mathrm{sgn}$ are functions that map to the nearest integer and to 1 (-1) for a positive (negative) signed real valued input respectively. We note that this representation is used for analysis, and is different from the state representation used as input for the RL agent.

This representation of a state allows us to map monomer positions $\{\mathbf{r}\}$ to a state representation that in turn maps to an unique identifier. We can also group states together based on two operations: inversion and chirality. Inversion consist of inverting the ordering of the monomer labeling in the chain, \textit{i.e.}, $1 \mapsto N$, $2\mapsto N-1$, and so on. Chirality consists of flipping the backbone angles, \textit{i.e.}, $\phi_{i-1, i, i+1} \mapsto 2 \pi - \phi_{i-1, i, i+1}$. We note that for odd length chains with two alternating flavors, \textit{e.g.} BYBYBYB, the system is fully symmetric under inversion. For such chains of even length the system requires a switching of flavors, \textit{i.e.} B$\mapsto$Y and Y$\mapsto$B, to be symmetric. When we group states based on these symmetries we explicitly mention it in the text.

We define geometries to be rigid bond arrangements, \textit{i.e.}, no zero energy deformations of the monomer positions are possible, that describe a unique 2D spatial arrangement of the monomers. We classify a state's geometry by mapping the adjacency matrix of bonds to a canonical representation per geometry that is invariant under all monomer index permutations. Folds are distinct backbone arrangements within a geometry. 

\subsubsection{Repairing incomplete pathways}
When we sample pathways in the test phase, sometimes multiple bonds are made since the last snapshot of the simulation. This is due to the simulation time taken before returning a snapshot when a new bond is detected: to ensure that the bond is sufficiently stable we wait one unit of simulation time before recording a snapshot. Within this time window, it is possible for the system to form or break another bond. 

To determine the strategy type of a pathway and the types of states it traverses, we require the full pathway through the tree, meaning that each transition between states corresponds to a single bond forming or breaking (in the case an interaction type is turned off, it is possible that several bonds break simultaneously). Using a full tree of states and their possible transitions, we are able to fill in the gaps of sampled pathways. Because multiple pathways might exist between two recorded states, we record all these possible pathways to create a list of new full pathways. To prevent overcounting of certain pathways, full pathways are always linked to their original, possibly broken, pathway and are reweighted according to the number of new pathways relating to the original pathway in statistical analysis. 

\subsubsection{Pathway analysis}
We consider a sequence of indices as a path $\mathcal{P}$ through the folding tree. The length of this path $|\mathcal{P}|$ is equal to the total number of states visited. These states are categorized as either transient (T), kinetic trap (KT) or global minimum (GM) as described in the Main Text. In turn, pathways can be grouped by if they cross a GM, if they cross a KT, or if all states are of type T, where the first and last state are excluded because they always are GM. Pathways that only cross GM are thermodynamically favored: the system always converges towards GM over sufficiently long time. Pathways that only cross KT or T states are typically not thermodynamically favored, but might be kinetically favored.

The length of the path allows us to differentiate between functionally different pathways of folding. We first differentiate between pathways with and without backtracks. A path contains a backtrack if one or multiple states are visited before returning to an earlier visited state and continuing down another set of states. The set of states from the first encounter of this earlier visited state up to the state before visiting this state again is a backtrack. A path with backtracks removed that is as long as the shortest distance between the start and end states is a downhill path: along the effective path bonds are only formed and never broken. In contrast, a path with backtracks removed that is longer than the shortest path between the start and end point is a detour path. Note that paths are either downhill or detour and both can have backtracks. We note that removing backtracks is loop erasure and the result is a self-avoiding path.

\subsubsection{Action graphs}
These pathways can be grouped into distinct folding strategies. Inspired by actuation graphs to analyze transition paths in hysteretic systems~\cite{meulblok2026path}, we plot our pathways in \textit{action graphs}: a visual representation of the folding tree grouped by active secondary interactions. 
Each combination of active secondary interactions is represented by a binary string of length 3 $(b_{BB}, b_{BY}, b_{YY})$, where $b_x$ is 1 (0) if interaction $xy$ is on (off). Each combination groups all states where these interactions are (in)active together in the \textit{interaction region}.
Within a region, the system can transition to other states within this space without the need to take another action, \textit{i.e.} turn on or off an additional interaction. Conversely, to transfer to another region the system needs to turn off (on) an (in)active interaction. Transitions within a region are represented by black arrows and transitions to another region by colored arrows. 

We group states in an interaction region together by their number and type of secondary interactions. The reason for this is threefold. First, this improves visual clarity. Second, the number of secondary interactions is indicative of how long the system should reside in any action space. Third, showing the secondary bond types allows us to distinguish in bond composition, which can be targeted during folding by tuning the relative interaction strengths. 
States within an action space are grouped by normalized barycentric coordinates for the triplet $(n_{BB}, n_{BY}, n_{YY})$ with respect to an equilateral triangle (Fig.~\ref{fig4}A), where $n_I$ is the the number of secondary bonds of type $I$ in the state. To prevent overlap, these coordinates are offset by the total number of secondary bonds $n$ in the vertical direction. This allows us to see at a glance the bond type distribution for any state in the pathway.

We define a motif as a unique sequence of interactions $m = (I_1, I_2, \ldots, I_k)$ of length $k$, obtained by reducing the raw action sequence $(a_1, a_2, \ldots, a_t)$ of a run in the test pathways. Because the agent acts at every folding step, it can repeatedly select the same interaction to let the chain continue folding within one interaction region, so the raw sequence contains strings of actions on a single interaction type that together correspond to a single experimental step. We therefore partition the sequence into maximal blocks of consecutive actions acting on the same interaction type, and reduce each block according to its net effect: a block that leaves the interaction in a different state than it started in becomes a single entry $I_i$, whereas a block that returns it to its initial state---a back-and-forth in which bonds of that type are broken and re-formed---becomes two consecutive entries of that type. Repeated switches within a block that cancel out are discarded as retries. 
For example, the downhill strategy of Fig.~\ref{fig4}B1 is the motif (BY, YY, BB), and the detour strategy of Fig.~\ref{fig4}B2 is the motif (BY, YY, BY, BY, BB). Note that the direction of each switch follows from the ordering as the chain starts with all secondary interactions off.
The reduction collapses waiting and retries while retaining every switch that changes active interactions. Note that a single motif may be realized by pathways that do and do not backtrack. We apply this reduction to every run in the
test pathways and build a collection of motifs per target geometry.

\subsubsection{Mapping motifs to experimental protocols}
A motif specifies a protocol once we fix three things: the order of switches, the waiting time between them, and the strength at which each interaction is turned on. The order follows directly from the motif: $m = (BY, YY, BY, BY, BB)$ (Fig.~\ref{fig4}B2) is the protocol BY on, YY on, BY off, BY on, BB on. The waiting times we estimate from the test pathways that make up the motif. States within an interaction region are ordered by their number of secondary bonds $n$ (Fig.~\ref{fig4}A), so $n$ tracks how far a chain has progressed through the region. We measure the fraction of chains at each $n$ as a function of simulation time, and set each waiting time to the time where the occupancy peaks at the value of $n$ at which the motif's next switch occurs. The aim is to reach the largest fraction of the population of folding chains where the switch is meant to act. Because the test pathways immediately continue folding as the agent dictates the subsequent interaction to switch, this method crudely estimates the most probable time to reach that $n$. Finally, the strength: where a motif requires the chain to reach the global minimum of an interaction region, we turn the interaction on at a reversible strength, $\epsilon_{I} \in [4,12]$, and wait longer, so that bonds can break and re-form and the chain converges to the thermodynamically favored state instead of trapping. Backtracking pathways can also be actuated with reversible interaction strength (Fig.~\ref{fig5}A2).

\newpage

\section{Supplementary Information}
Here we provide results complementary to those shown in the Main Text. 

\subsection{State space enumeration}
For small chain lengths $(N \leq 9)$, state space can be enumerated in full on a computer. Tab.~\ref{tab:statespace} shows the number of states and transitions per $N$. We note that this space contains only the information of what states are possible and which are a single bond addition or removal removed from one another. It contains no information of flavor sequence, secondary interactions, or kinetic rates.

\begin{table}[b]
    \caption{Number of states and transitions between states per chain length $N$.}
    \centering
    \begin{tabular}{l r r r}
       $\mathbf{N}$  & \textbf{no. states} & \textbf{no. transitions} & \textbf{no. geometries}\\\hline
       6 & 161 & 900 & 3 \\
       7 & 1028 & 7360 & 4\\
       8 & 6646 & 57842 & 9\\
       9 & 42388 & 432878 & 16\\
    \end{tabular}
    
    \label{tab:statespace}
\end{table}
\subsection{Training results}
\begin{figure}[h]
    \centering
    \includegraphics[]{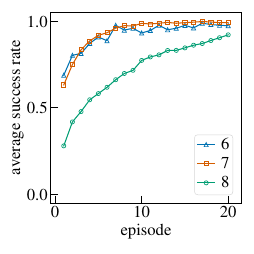}
    \caption{Fraction of successful pathways that reach the intended target geometry per training epoch of the RL agent. Success rate is averaged over all possible target geometries for chains of length $N=6, 7$ and $8$ (legend).}
    \label{fig:RL_r1}
\end{figure}

Our training results for chains of lengths $N=6, 7, 8$ are shown in Fig.~\ref{fig:RL_r1}. 

\subsection{Test results}
\begin{figure*}[h]
    \centering
    \includegraphics[width=1\linewidth]{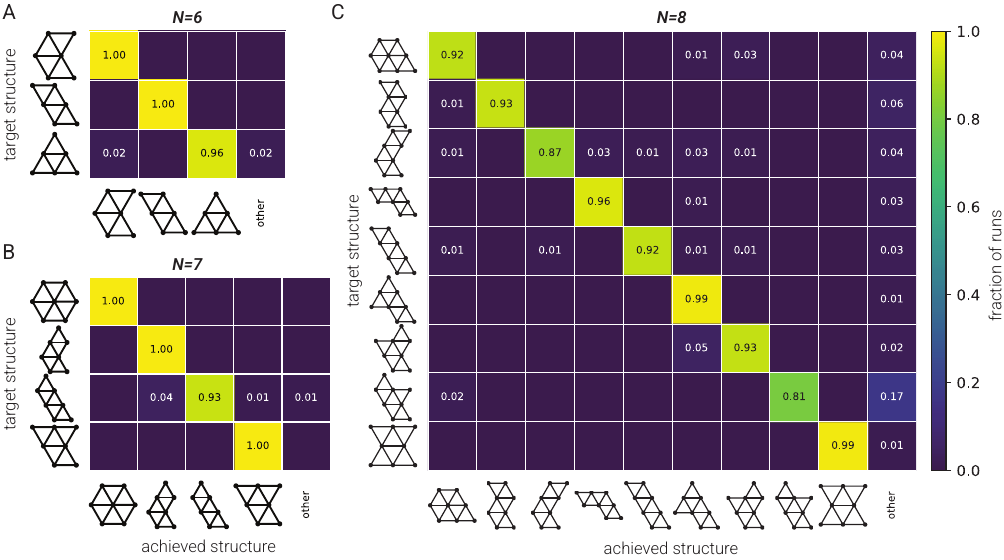}
    \caption{Pathway end state distribution for test runs of trained agents for $N=6$ (A), $N=7$ (B) and $N=8$ (C), normalized per row. Pathways end when they reach the intended target or reach a maximum path length (14, 32 and 32 respectively). Rigid geometries are shown explicitly, other floppy states are grouped in the category ``other''.}
    \label{figSI:CM}
\end{figure*}

Test pathways are generated by taking the most likely action proposed by trained agents. The intended target geometry, and the geometry these pathways end up in is shown in Fig.~\ref{figSI:CM}. The vast majority of test pathways reach their intended target, with a small fraction of pathways failing to reach the target geometry within the allowed number of maximum folding steps (14 for $N=6$, 32 for $N=7$ and $8$). We note that the non-target geometries reached tend to be structurally similar to the intended target geometry, a consequence of branching pathways. 

\subsection{Distribution of pathway types}
\begin{figure*}[h]
    \centering
    \includegraphics[width=\textwidth]{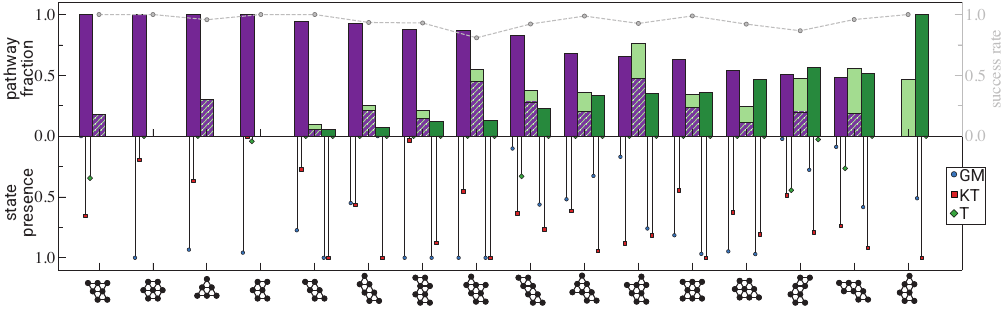}
    \caption{Top: fraction of pathway types (downhill: purple; detour: dark green) in 100 sampled test pathways per target geometry. Both pathway types can contain backtracking (light green) and the purple hash indicates the fraction coming from downhill. 
    Bottom: fraction of pathways that pass through a state of type GM (blue circle) or of type KT (red square), or through only states of type T (green diamond). Note that a single pathway may pass through states of multiple types, so the fractions need not add to 1.}
    \label{figSI:histogram}
\end{figure*}

The distribution of pathways types and the states they traverse is shown in Fig.~\ref{figSI:histogram} for all target geometries for chains of length $N=6$, $7$, and $8$, sampled from 100 test pathways per target.

\subsection{Ice-cream cone analysis}
Here we show how pathways sampled using the trained RL agent targeting the $N=8$ ice-cream cone geometry (Fig.~\ref{fig4}A) translate to different motifs which turn into protocols that sample different pathways in simulations.

\begin{figure*}[t]
    \centering
    \includegraphics[width=\textwidth]{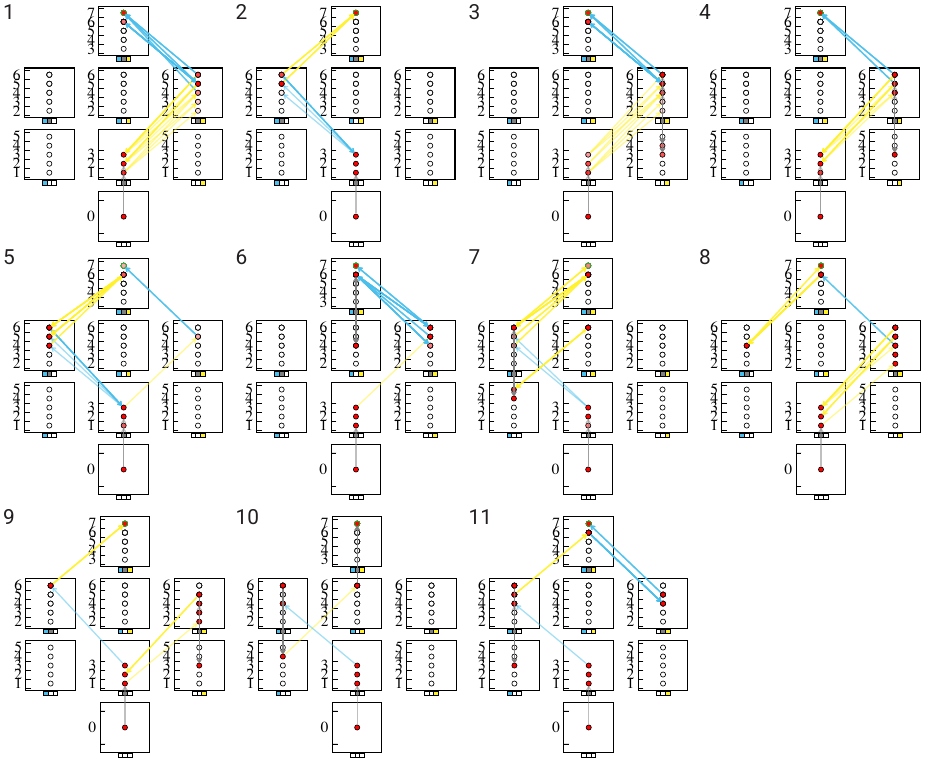}
    \caption{All motifs with successful pathways within the action graph of Fig.~\ref{fig4}A. Motifs are ordered by prevalence. Motifs 5, 7 and 8 show the YY off/on motif.}
    \label{figSI:motifs}
\end{figure*}

\begin{figure*}[h]
    \centering
    \includegraphics[width=\textwidth]{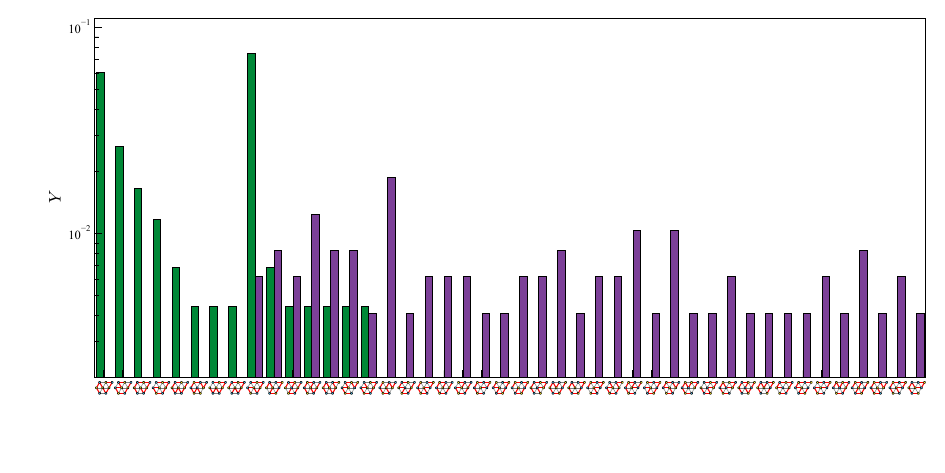}
    \caption{Histogram of the relative yield for the $N=8$ ice-cream cone geometry's folds for two protocols: downhill (Fig.~\ref{fig4}B1, purple) and detour (Fig.~\ref{fig4}B2, green). Backbone is shown in red, secondary bonds in gray.}
    \label{figSI:backbones}
\end{figure*}

\begin{figure}[h]
    \centering
    \includegraphics{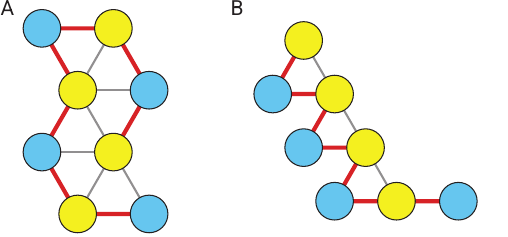}
    \caption{Two of the most visited states with positive selectivity in Fig.~\ref{fig4}C. The backbone is shown in red, secondary bonds in gray. Note that floppy states are grouped by adjacency matrix in Fig.~\ref{fig4}C, so the structure in B is a single possible backbone orientation out of multiple orientations that yield the same adjacency matrix.}
    \label{figSI:states}
\end{figure}

The action graph of the $N=8$ ice-cream cone (Fig.~\ref{fig4}A) consists of many motifs, shown in Fig.~\ref{figSI:motifs}. These motifs translate into protocols of switching interactions as described in the section on mapping motifs to experimental protocols. The protocols derived from motifs 1 and 3 are shown in Fig.~\ref{fig4}B. These protocols select different backbone arrangements within the $N=8$ ice-cream cone geometry. A full histogram of the relative backbone yields from 1000 simulations of individual chains folding per protocol is shown in Fig.~\ref{figSI:backbones}. This difference is a consequence of the two protocols actuating distinct pathways in state space (Fig.~\ref{fig4}C). The two most visited states in these pathways, excluding the chain, with a positive selectivity are shown in Fig.~\ref{figSI:states}. In the detour protocol, the s-shape geometry (Fig.~\ref{figSI:states}A) is reached before breaking any bonds, while the structure in Fig.~\ref{figSI:states}B is reached after the breaking step.

\begin{figure*}[t]
    \centering
    \includegraphics[width=\linewidth]{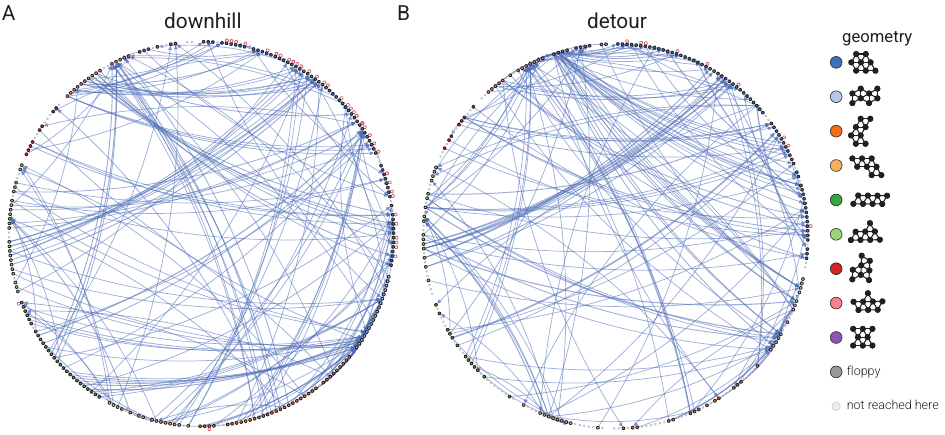}
    \caption{Graphs showing the transitions between states (nodes) from before and after the YY off/on motif was applied to the downhill (A) and detour (B) protocols of Fig.~\ref{fig4}B. The width of the edges correspond to how often this transition occurred. Transitions between different states are marked by blue arrows, transitions from a state to itself are marked by red loops.}
    \label{figSI:graphs}
\end{figure*}

In Fig.~\ref{fig4}D, the distributions of folded geometries are generated from 10 independent bulk simulations of 50 chains folding subject to either the downhill or detour protocol. Chain configurations are grouped into states, allowing us to count how many states belonging to a geometry there are at the end of the protocol (Fig.~\ref{fig4}D1) and after a YY off/on motif is appended (Fig.~\ref{fig4}D2). A more detailed picture of this transition between distributions is shown in Fig.~\ref{figSI:graphs}. There, transitions between states from the end of the protocol to the end after the YY off/on motif are shown explicitly, with states represented as nodes and transitions as directed edges. This YY off/on motif presented itself in motifs 5, 7 and 8 (Fig.~\ref{figSI:motifs}).

\begin{figure}
    \centering
    \includegraphics{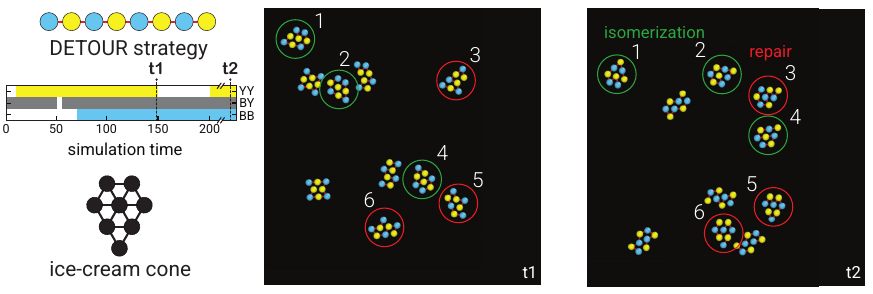}
    \caption{Two simulation snapshots of a bulk simulation of ten $N=8$ chains after folding under the detour protocol at time t1 and t2. Ice-cream cone structures isomerize (green circles), structures with other geometries can repair (red circles). The remaining structures reconfigure into the s-shape geometry.}
    \label{figSI:repair}
\end{figure}

The YY off/on switch redistributes the ensemble of structures mostly towards the ice-cream cone and s-shape geometries (Fig.~\ref{fig4}D2). In a bulk simulation of 10 chains, we capture this redistribution and identify two mechanisms: isomerization, where ice-cream cone structures retain their geometry but change color arrangement; and repair, where non ice-cream cone configurations reconfigure into the ice-cream cone geometry (Fig.~\ref{figSI:repair}). 

\begin{figure*}[t]
    \centering
    \includegraphics[width=\linewidth]{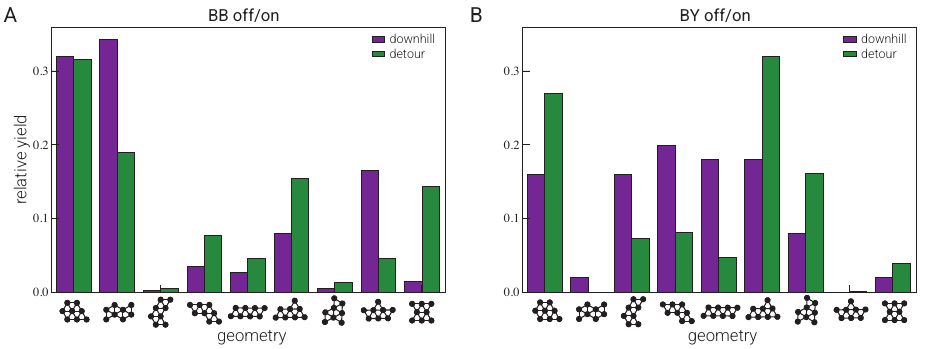}
    \caption{Distribution over geometry after applying a BB off/on motif (A) or BY off/on motif (B) to the folds produced under the downhill (purple) and detour (green) protocols in Fig.~\ref{fig4}D1. }
    \label{figSI:histogram_switches}
\end{figure*}

The interaction type in the off/on motif changes the distribution of geometries after the protocol (Fig.~\ref{figSI:histogram_switches}). For the BB off/on motif the distribution remains largely the same to the initial distribution. For the BY off/on motif the s-shape (no.~2) and crown (no.~7) geometries are suppressed and the distribution is spread over the remaining geometries.

\subsection{Supramolecular assembly of the pencil geometry}
\begin{figure*}[h]
    \centering
    \includegraphics{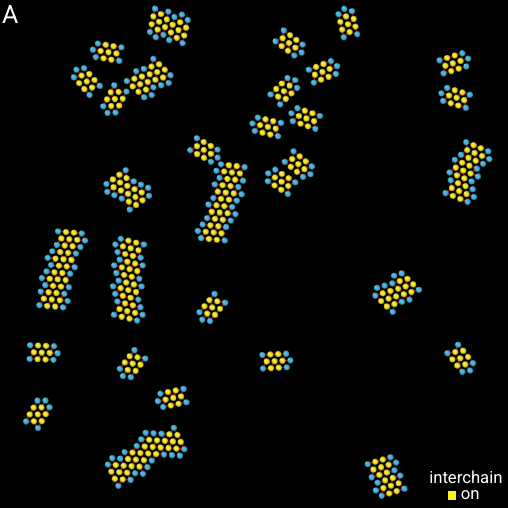}
    \caption{Simulation snapshot of a bulk simulation of fifty $N=12$ chains each folded into the same backbone arrangement of the pencil geometry with all intrachain interactions on. The structures associate into extended ribbons when interchain YY interactions are turned on.}
    \label{figSI:supra_pencil}
\end{figure*}

The supramolecular assembly of the pencil geometry (Fig.~\ref{fig5}B) yields uniform, ribbon-like structures, Fig.~\ref{figSI:supra_pencil}.

\newpage

\section{Supplementary Videos}
Here we provide descriptions for the videos accompanying this work.

\noindent\textbf{Supplementary Video 1 \textbar\ Downhill and detour folding of $N=8$ alternating chains into the ice-cream cone geometry.} Two separate simulation boxes each with fifty chains folding under the downhill (left) and detour (right) protocols of Fig.~\ref{fig4}B, followed by the YY off/on motif. The middle plot shows the selectivity of the states in each timestep, sized by occupancy. Counters on top display the number of ice-cream cone geometries. Active intrachain interactions shown in the bottom corners.

\noindent \textbf{Supplementary Video 2 \textbar\ Detour folding of $N=7$ alternating chains into geometries with binding pockets.} Fifty chains folding under the detour protocol of Fig.~\ref{fig5}A1. Active intrachain interactions shown in the bottom left.

\noindent \textbf{Supplementary Video 3 \textbar\ Backtrack folding of $N=7$ alternating chains into hinging structures.} Fifty chains folding under the backtrack protocol of Fig.~\ref{fig5}A2. Active intrachain interactions shown in the bottom left.

\noindent \textbf{Supplementary Video 4 \textbar\ Downhill folding of $N=7$ alternating chains into the flower geometry.} Fifty chains folding under the downhill protocol of Fig.~\ref{fig5}A3. Active intrachain interactions shown in the bottom left.

\noindent \textbf{Supplementary Video 5 \textbar\ Downhill folding of $N=12$ chains into the pencil geometry.} Ten chains with sequence BYBYYBYBYYYB folding under the downhill protocol of Fig.~\ref{fig5}B. Active interactions shown in the bottom left.

\noindent \textbf{Supplementary Video 6 \textbar\ Supramolecular assembly of $N=12$ structures folded under a downhill protocol.} Fifty folded chains assembling under YY interchain interactions. The initial folded structures are generated by 5 separate bulk simulations of 10 chains folding under the downhill strategy of Fig.~\ref{fig5}B. Active interchain interactions shown in the bottom left. Playback speed shown in top left.

\noindent \textbf{Supplementary Video 7 \textbar\ Downhill folding of $N=9$ chains into the C-shape geometry.} Ten chains with sequence BYYYYYYYB folding under the downhill protocol of Fig.~\ref{fig5}C. Active interactions shown in the bottom left.

\noindent \textbf{Supplementary Video 8 \textbar\ Capture and release of a guest particle by a $N=9$ reconfigurable structure.} A single folded C-shape structure captures a guest particle (red) and releases it by refolding into an inert shape. Smaller black particles are the solvent. Active interactions shown in the bottom left. Playback speed shown in the top left.

\end{document}